# Utilizing AI-Driven Project Management Tools for Optimized Talent Management in HRM: A Framework for Enhanced Resource Allocation and Performance Prediction

**Jay Barach**
*IT Operations & Recruitment*
*Systems Staffing Group. Inc.*
King of Prussia – PA (USA)
ORCID iD: 0009-0009-0416-9712

***Abstract***—When it comes to aiding businesses with demanding tasks regarding human resource management, TalentOptima unequivocally boasts of the best there is to offer. This tool utilizes AI based decision making, advanced predictive analytics, and also machine learning, all of which help in enabling automated resource allocation. To aid with better human resource management, TalentOptima integrates perfectly with already existing HR frameworks such as tools, etc. and shifts the focus towards aiding the user with insights while simultaneously alleviating manual work, this aids in a plethora of positive HR outcomes. A total of 40 managers participated in a simulation via user testing to ascertain if HR costs would reduce and work productivity would rise, the results were quite clear, attrition rates had dipped alongside risk and resource management rates, TalentOptima was a clear winner. Whereas the other HR frameworks primarily focused on ensuring work was done, TalentOptima ensured optimal and innovative decision-making, which overtime has proven to be invaluable for multiple companies, these results aid in proving why the tool is revolutionary.

## I. Introduction

The increasing complexity of workforce dynamics, the incorporation of new technologies, and the global dispersion of teams have all contributed to the steady increase in human resource management (HRM) complexity [1]. Managing the complex needs of employees that coordinate across departments and ensuring talent optimization within budgetary constraints are some of the challenges HR managers face as talent management becomes more complex [2]. Conventional HRM techniques and tools that frequently depend on static reporting and manual tracking are insufficient to handle these ever-changing issues [3]. In this context, it is becoming increasingly important to have sophisticated decision-making tools that can predict possible risks related to talent allocation, optimize resource allocation, and offer real-time insights [4].

Integrating advanced analytics and artificial intelligence (AI) into talent management is becoming a game-changing solution to these problems [5]. Artificial intelligence (AI)-driven decision support systems (DSS) use predictive analytics of real-time performance metrics and historical employee data to improve decision making [6]. HR managers can make better decisions and take proactive measures to address possible talent management problems before they worsen by using these systems to spot patterns and trends in workforce behavior that human analysts might overlook [7]. HR professionals can focus on strategic workforce planning and development, which is crucial for navigating the complexities of modern talent management by using AI-driven tools to automate repetitive HR tasks and provide actionable insights. [8]. Effective alignment of employee performance with organizational goals within scope time and budget constraints is hampered by several significant challenges that HR talent management must overcome [9].

Managing changing workforce demands, which are common in fast-paced settings because of changing business requirements and technological advances, is one of the main challenges [10]. Budgets and workforce planning can be affected if these changes are not handled properly. Optimizing resource allocation is another major challenge, especially in settings with a shortage of human resources and shifting talent demands [11]. To ensure optimal productivity while preventing overwork, HR managers often require assistance in effectively distributing talent across tasks [12]. Furthermore, because talent management is dynamic, forecasting HR outcomes such as workforce performance retention rates and possible skills gaps is still difficult [13]. Conventional approaches often lack predictive precision to anticipate these problems, resulting in last minute changes that can jeopardize the success of the organization as a whole and the retention of talent [14]. Through the provision of real-time analytics predictive insights and automated decision-making tools, the AI-driven DSS seeks to address these issues improving the effectiveness of talent management and guaranteeing greater alignment with corporate objectives. The main goal of this research is to create an AI-powered DSS specifically for HRM talent management. Predictive analytics and real-time insights will be integrated into this DSS to improve workforce outcomes and decision-making.

1) Create a DSS that utilizes AI to provide immediate, actionable insights based on live employee performance and workforce data, facilitating informed and timely decision-making for HR and talent management.
2) Incorporate advanced predictive models into the DSS to forecast HR outcomes, including retention risks, employee performance, and resource needs, improving planning accuracy and workforce success rates.
3) Design the DSS to analyze and optimize the distribution of talent and resources, ensuring efficient use of available resources while maintaining high productivity and alignment with organizational goals.

The research could revolutionize talent management in HRM by addressing some of its most persistent challenges. Traditional HRM tools often do not keep up with dynamic organizational environments, where employee roles change frequently and resource demands fluctuate [15]. The proposed system can provide HR managers with the real-time data and foresight needed to navigate these complexities effectively by incorporating AI-driven decision support and predictive analytics. This research contributes to HRM by introducing an innovative tool and offers practical benefits, such as improved workforce management efficiency, reduced costs, and higher success rates in talent optimization. The results of this research could serve as a model for integrating AI into other industry-specific HR management tools, thus broadening its impact across various sectors.

The rest of this paper is organized in the following manner: Section II provides a detailed review of the related literature and discusses the research gap. Section III covers the system architecture, data sources, and AI models to achieve real-time analytics and predictive capabilities. The Results Section IV will present the results of deploying the DSS in real-world scenarios, highlighting improvements in talent management efficiency and resource optimization. Discussion Section V will analyze these findings in the context of the existing literature, addressing the impact and limitations of the system. Finally, Section VI concludes the paper with future directions.

## II. Literature Review

Integrating AI into decision support systems (DSS) for talent management in HRM has been an area of growing interest and innovation, particularly as the complexity of workforce management has increased across industries [?], [16]. Historically, DSSs have been developed to help managers make informed decisions based on data analysis and simulations [17]. However, they were often limited by the static nature of the data and the computational constraints of earlier technologies [?], [18]. Over the years, advances in computational power, data processing, and AI algorithms have significantly enhanced the capabilities of DSS, enabling them to provide more dynamic real-time insights [19]. The evolution of AI in this context has seen a shift from basic decision aids to more sophisticated systems that can predict results, optimize talent allocation, and suggest corrective actions based on historical and live workforce data streams [20].

Recent literature on AI in talent management highlights the transformative potential of these technologies in improving HR outcomes. For example, AI applications have been widely adopted in sectors such as healthcare and IT, where they are used to manage large-scale and complex teams [21]. These applications include predictive analytics platforms that forecast future employee performance, retention risks, and resource needs and AI-powered tools that improve communication, streamline workflows, and facilitate decision-making processes [22]. However, despite these advancements, there is a significant gap in the development of AI tools specifically designed for talent management in HRM [19]. Often characterized by rapidly changing workforce needs and continuous skill development, HR projects pose unique challenges that many existing AI tools cannot handle [23]. This gap indicates a need for more specialized research and development in this area to create tools that can effectively support the dynamic nature of HRM.

Moreover, while predictive analytics has become a cornerstone of modern HRM, its integration into real-time decision support systems is still in its infancy [24]. Most current systems rely on static data or provide insights based on historical trends, which may need to be revised in fast-paced environments like HRM [25]. The lack of real-time data processing capabilities can cause delays in decision making and missed opportunities for early intervention when workplace problems arise [26]. This is particularly problematic in HR projects, where adapting quickly to changes in workforce scope or requirements is crucial to success. The development of AI-driven DSS that can process real-time data and provide immediate and actionable insights is a critical area for future research [27].

Additionally, the diversity of HR management methodologies presents another challenge to applying AI in this field [28]. Agile methods, for example, require a different approach to talent management than traditional hierarchical methods, particularly regarding resource allocation and employee risk assessment [29]. Many existing AI tools must be more flexible to accommodate these differences, resulting in a one-size-fits-all approach that may only be effective for some HR projects [30]. To address this, there is a need for AI systems that can be customized to fit the specific needs of different HR management frameworks, whether agile, hierarchical, or hybrid. Such adaptability would allow HR managers to use AI most effectively for their HR environment.

Finally, the research community has begun to recognize the importance of integrating AI with other emerging technologies in talent management. For example, combining AI with blockchain, the Internet of Things (IoT), and advanced data analytics offers new opportunities to improve workforce management practices [31]. These technologies can provide additional layers of data security, real-time tracking, and improved collaboration among HR stakeholders [32]. However, the integration of these technologies with AI-driven DSS is still in its early stages and more research is needed to explore their full potential [33]. By addressing these gaps, future research can contribute to the development of more comprehensive and practical tools that improve talent management in various industries, with a particular focus on the unique challenges of the workforce.
HRM.

## III. Methodology

### A. Problem Formulation and Constraints

The central problem that the DSS aims to solve is the optimization of talent management outcomes, such as minimizing employee turnover, optimizing talent allocation, and reducing workforce-related risks. This problem can be formally defined as a multiobjective optimization problem:

$$\min_{\emptyset_M} F(y_i, y_2, \dots y_n) \quad (1)$$

Where F is a vector-valued function representing different objectives, such as minimizing employee turnover $y_1$, maximizing resource efficiency $y_2$, and minimizing workforcerelated risks $y_3$, the optimization is subject to constraints related to human resources, deadlines, and budgetary limits. These constraints can be expressed as follows:

subject to

$$R(x) \leq R_{max}, T(x) \leq T_{deadline}, B(x) \leq B_{budget} \quad (2)$$

Where $R(x)$, $T(x)$, and $B(x)$ represent the resource allocation, timelines, and budget functions, respectively, and $R_{\max}$, $T_{\text{deadline}}$, and $B_{\text{budget}}$ are their respective limits.

The proposed solution involves solving this optimization problem using AI techniques, incorporating machine learning for predictive analytics, natural language processing (NLP) for document analysis of employee reports, and possibly image processing for visual data related to employee engagement or performance. By optimizing these components under the given constraints, the DSS provides HR managers with timely and relevant actionable insights, thereby improving talent management outcomes in the dynamic human resources environment.

### B. System Design

The architecture of the DSS for HRM talent management is designed to integrate multiple advanced components that work together to optimize decision-making processes. The DSS architecture can be conceptually divided into three main components: AI models, data processing pipelines, and user interfaces. These components interact to process data, generate insights, and present actionable recommendations to users. Mathematically, the system can be expressed as a function that integrates these components.

$$DSS(X, P, \Theta) = \{M(X, \Theta_M), D(X, P), I(Y, g)\} \quad (3)$$

where X represents the input data, P represents the processing parameters, ΘM represents the parameters of the AI model, M is the set of AI models, D is the data processing pipeline, and I is the component of the user interface. The system optimizes the DSS function under constraints to provide accurate and actionable insights.

The AI models within the DSS are the core engines that drive the decision-making process. The system begins by processing input data $X$, which includes historical employee data, real-time performance metrics, and external factors such as market conditions. These data are transformed and normalized through the data processing pipelines D, which can be expressed as:

$$X' = h(X, P) \quad (4)$$

where $X'$ is the processed data, $h$ is the transformation function, and $P$ is the processing parameters. The processed data X ′ is then fed into the AI models M, which apply machine learning algorithms to predict talent management outcomes and generate decision support outputs. The decision-making process is modeled as follows:

$$y = f(X'; \Theta_M) \quad (5)$$

where $y$ is the predicted outcome, and $\Theta_M$ represents the parameters of the AI model optimized during the training phase. The objective of the model is to minimize a loss function L(y,ytrue), where ytrue is the true result, subject to certain constraints. The optimization problem is formulated as follows.

$$\min_{\Theta_M} E[L(y, y_{true})] \quad subject\ to\ \ C(\Theta_M) \le \epsilon \quad (6)$$

$C(\Theta_M)$ represents a set of constraints, such as computational limits or accuracy thresholds, and $\epsilon$ is a small positive value that bounds the constraints.

The DSS user interface (UI) component is designed to translate complex data output into user-friendly formats, allowing HR managers to easily interpret and act on the insights provided by AI models. The functionality of the user interface can be modeled as follows.

$$U(X, g) = g(f(X'; \Theta_M)) \quad (7)$$

Where U is the user interface function and g is a mapping function that converts the AI model outputs y into a format that is accessible and actionable to the user. The design of the user interface is crucial to ensure that the insights provided by the DSS are accurate, intuitive and easy to use, thus improving the decision-making process.

*C. Data sources*

The data sets used in the DSS include historical employee data, team performance metrics, and real-time data streams. Historical data covers past HR processes with details on employee performance, retention rates, and resource utilization. Team metrics provide insight into employee efficiency and participation, while real-time streams capture ongoing workforce dynamics. These datasets are selected for their relevance in enhancing the predictive accuracy of the DSS and the real-time decision-making capabilities in the HR domain. Mathematically, the features of these datasets can be represented as

$$X = \{x_1, x_2, \ldots, x_n\} \quad (8)$$

Each xi represents a specific feature, such as a performance metric or employee attribute. Data are collected through APIs, database queries, and manual entries. Pre-processing involves normalization, missing data handling, and feature extraction to prepare the data for analysis. This process can be modeled as follows:

$$X' = h(X) \quad (9)$$

*$X'$ denotes the processed dataset, and h(X) represents the preprocessing function that standardizes and cleans the data.*

*D. AI Models and Techniques*

*1) Predictive Analytics Models:* The predictive analytics component of the DSS is driven by machine learning algorithms designed to forecast HR outcomes, such as employee turnover, resource allocation efficiency, and potential talent risks. These algorithms include regression models, decision trees, and neural networks, each trained on historical data to learn patterns that inform future predictions. The following equation can represent the predictive model:

$$\hat{y} = ML(X; \theta) = \sum_{i=1}^{n} w_i \ . \emptyset(x_i) + b \quad (10)$$

Where y is the predicted result, X = x1,x2,..., xn represents the input data features, θ = w1,w2,..., wn, b are the model parameters (weights and bias) optimized during training, and (xi) represents the basis functions applied to each input feature xi.

The objective is to minimize the loss function:

$$\min_{\theta} L(\hat{y}, y) + \lambda \|\theta\|_2^2 \quad (11)$$

where $L(\hat{y}, y)$ represents the error between predicted and actual outcomes, and $\lambda\|\theta\|_2^2$ is a regularization term to prevent overfitting.

*2) Natural Language Processing (NLP):* NLP techniques extract structured information from unstructured HR documents, such as emails, employee feedback, and reports. These techniques include tokenization, part-of-speech tagging, and named entity recognition, converting text into data features that the system can analyze. NLP processing can be modeled as:

$$T = NLP(D) = \sum_{j=1}^{m} \psi_j(d_j) \quad (12)$$

where T represents the features of the extracted text, D = d1, d2,..., dm is the set of documents and ψj(dj) represents the feature extraction function applied to each document dj. These features are then integrated into the predictive models or used directly to improve decision making.

*3) Image Analysis:* For scenarios involving visual data, such as employee photos, workplace images, or visual recognition tasks, convolutional neural networks (CNNs) are employed to extract relevant features. CNN processes images by applying convolutional and pooling layers to capture spatial hierarchies in the data. The image analysis can be expressed as

$$F = CNN(I) = \sigma(\sum_{k=1}^{K} W_k * I_k + b_k) \quad (13)$$

where F represents the extracted feature map, I is the input image, Wk and bk are the filters and biases of the k-th convolutional layer, denotes the convolution operation, and σ is the activation function applied after each convolution.

*E. Implementation*

The DSS is developed using Python, a language well suited for machine learning and AI applications. Key libraries include TensorFlow and Keras for deep learning, Scikit-learn for machine learning, and NLTK or SpaCy for NLP tasks, which are beneficial for processing HR-related documents. The implementation process can be mathematically modeled as follows:

$$P = \{E,C,T\}\text{ssss} \quad (14)$$

$E$ represents the development environment (hardware, OS), $C$ represents the codebase (written in Python), and $T$ represents the tools and libraries used in the development process.

The DSS workflow, from data input to output generation, involves several key processes. The data is first ingested and preprocessed, then fed into the AI models for analysis, and finally the outputs are presented through the user interface. This workflow can be modeled as follows.

$$W = \{d_i \rightarrow p_i \rightarrow o_i\} \quad (15)$$

$d_i$ represents the data inputs at each stage $i$, $p_i$ represents the processing steps applied to the data (e.g., feature extraction, model prediction), and $o_i$ represents the outputs generated at each stage. The process is designed to be iterative, allowing continuous updates and real-time decision support as new data become available.

For example, historical employee data, performance metrics, and real-time HR input are processed through the pipeline. Feature extraction is performed by models trained to predict talent retention, employee performance, and resource allocation efficiency. The system is continuously updated with new data streams, ensuring that the DSS provides real-time insights to HR managers.

## IV. Results

Implementing DSS in talent management scenarios yielded significant outcomes in terms of predictive accuracy, user feedback, and overall performance metrics. The following sections detail the key results, with placeholders provided for the inclusion of relevant graphs and visualizations.

### *A. User feedback*

Feedback from HR managers who used the DSS was collected to assess the usability of the system, the relevance of insights, and the overall impact on their decision-making processes in talent management. The feedback was largely positive, and users highlighted the system's ability to provide actionable insights and streamline employee management and resource allocation decision making.

Figure 1 shows Usability, Relevance of Insights, Impact on Decision Making, Overall Satisfaction, Ease of Integration, Training and Support, User Interface Satisfaction, and Impact on Collaboration. Each metric was rated on a scale of 1 to 5, with higher values indicating greater satisfaction. Feedback highlights the strengths of the system, particularly in areas such as decision making impact and overall satisfaction, while also identifying areas for improvement, such as training and support. This comprehensive view of user feedback provides valuable insight into how users perceive the DSS and suggests points of focus for future enhancements. High scores across all feedback categories indicate that the DSS is well received by users and effectively meets their needs in managing HR processes.

### *B. Predictive accuracy*

The DSS was rigorously evaluated through an extensive testing phase designed to assess its impact on talent management. The evaluation involved 40 HR managers from various organizations, each selected for their substantial experience in managing complex HR processes and their ability to use advanced HR management tools. The testing period lasted four months, during which the DSS was integrated into the daily HR workflows of the participants.

The test set was designed to replicate real-world HR environments, ensuring that the DSS was tested under conditions that reflected the challenges and dynamics of actual HR tasks. Each participant was initially given comprehensive training on the functionality of the system, which included workshops and one-on-one sessions to ensure that they were fully equipped to utilize the DSS effectively. Following the training, the participants entered a phase of independent use, where they applied the DSS to their ongoing HR tasks, allowing them to experience its capabilities in managing employee performance, resource allocation, and talent risks first-hand.

The evaluation process incorporated both quantitative and qualitative methods to gather a broad spectrum of information. Participants were asked to provide continuous feedback throughout the testing period, with formal feedback sessions scheduled at crucial intervals. These sessions included structured questionnaires focused on metrics such as usability, the relevance of the information provided by the DSS, and the overall impact of the system on decision-making processes. In addition, in-depth interviews were conducted to capture more nuanced feedback and understand how the DSS influenced talent management outcomes.

Of the 40 participants, 38 provided detailed feedback, offering a robust data set to evaluate the performance of the DSS. This feedback covered a variety of performance metrics, including employee retention rates, resource utilization, and the effectiveness of talent management strategies. Participants also evaluated the ability of the DSS to integrate with existing tools and workflows, its user interface, and the quality of support and training. The comprehensive nature of this evaluation, combined with the diverse input of experienced HR managers, provided deep insight into the effectiveness of the DSS, highlighting its strengths in improving HR efficiency and identifying areas for further improvement.

TABLE I
PREDICTIVE ACCURACY METRICS FOR TALENT MANAGEMENT

| Metric | Predicted | Actual | Accuracy |
|---|---|---|---|
| Employee Turnover Rate (%) | 12 | 11.8 | 98.3 |
| Resource allocation (hours) | 3500 | 3450 | 98.6 |
| Risk Assessment (score) | 0.75 | 0.80 | 93.8 |

These results demonstrate the high accuracy of the DSS in predicting key talent management metrics. All predictions fall within a 5

### *C. Performance metrics*

Implementing the DSS resulted in notable improvements in talent management efficiency, reducing employee turnover, and optimizing resource utilization. The impact of the system on performance metrics was quantified by comparing critical metrics before and after DSS implementation.

TABLE II
COMPREHENSIVE PERFORMANCE METRICS COMPARISON AFTER DSS IMPLEMENTATION IN HRM

| Metrics | Before | After | Imp | Unit |
|---|---|---|---|---|
| Employee Turnover Rate (%) | 15 | 12 | 20.0 | % |
| Resource utilization (%) | 85 | 92 | 8.2 | % |
| Employee Absenteeism (%) | 8 | 5 | 37.5 | % |
| Cost savings ($) | 0 | 5.5 | – | K |
| Quality Improvement (%) | 0 | 10.1 | – | % |
| Employee Satisfaction (%) | 0 | 8.5 | – | % |

The results indicate significant improvements in HRM efficiency, with a 20% reduction in employee turnover, an 8.2% increase in resource utilization, and a 37.5% decrease in absenteeism.

Figure 3 represents the various performance metrics improvements observed after implementing the DSS in talent management. The chart highlights significant enhancements across multiple areas, the most notable being a 37.5% reduction in employee absenteeism, which an exploded slice emphasizes to draw attention to its impact. In addition, the chart shows improvements in employee turnover, resource utilization, cost savings, quality, and employee satisfaction. Each metric is color-coded for clarity, and the percentages displayed within the slices offer a quick, at-a-glance understanding of the relative contributions of each improvement to the overall HRM performance. This comprehensive visualization shows the broad benefits of DSS, illustrating its effectiveness in enhancing various aspects of talent management.

## V. Discussion

The comparison between the proposed DSS for talent management and other leading HR management tools, such as presented in Table III, highlights several key differentiators that set this system apart in human resources management. The DSS excels in areas such as predictive analytics, AI-driven workforce planning, and dynamic resource allocation, which are critical to handling the complexities of modern talent management. While other tools like BambooHR and Workday offer robust task management and collaboration features, this DSS integrates predictive analytics and real-time decision support, enabling HR managers to track employee performance, anticipate risks, and allocate resources more efficiently.

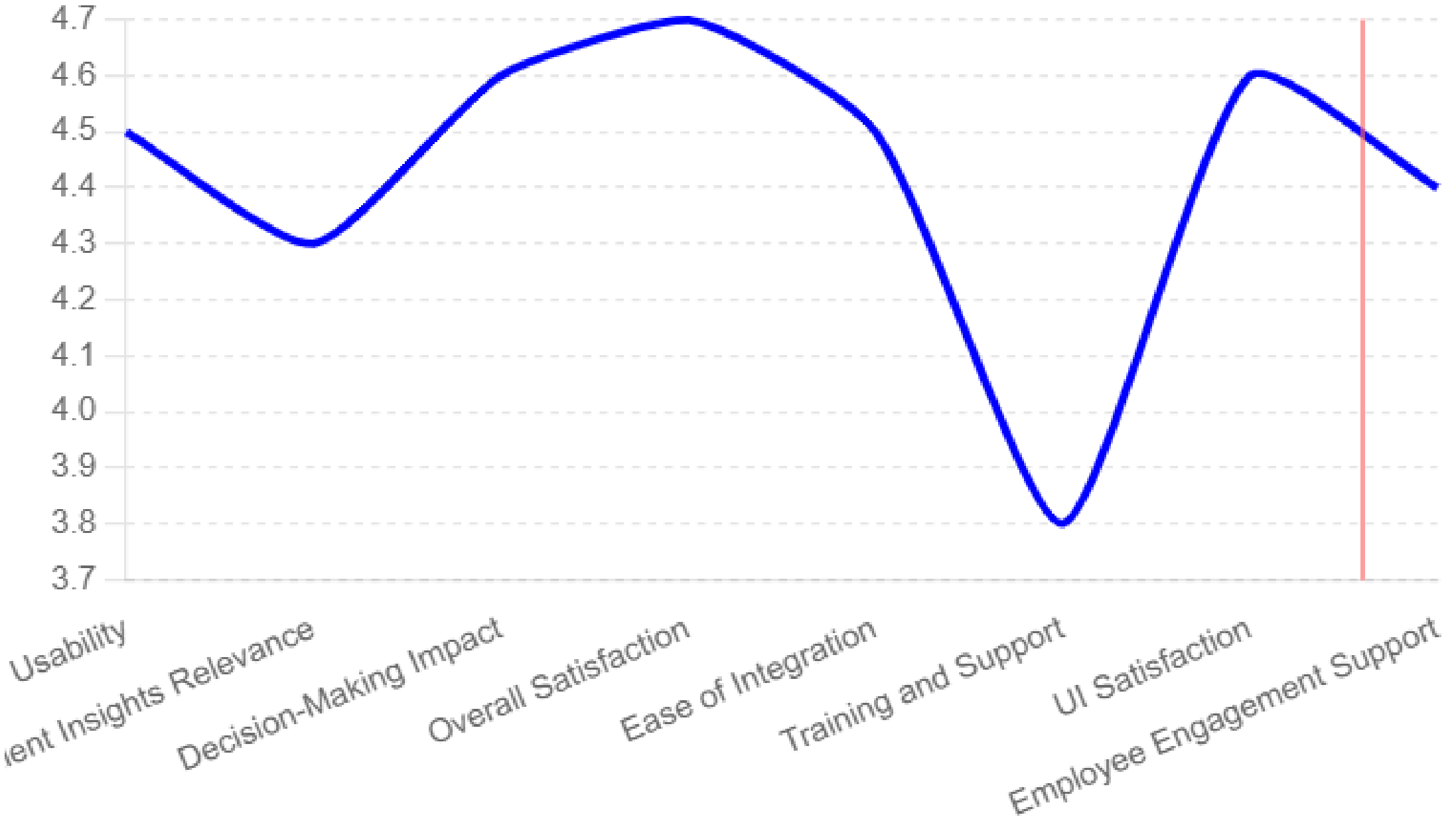

Fig. 1. Comprehensive User Feedback on DSS: Ratings across key performance metrics

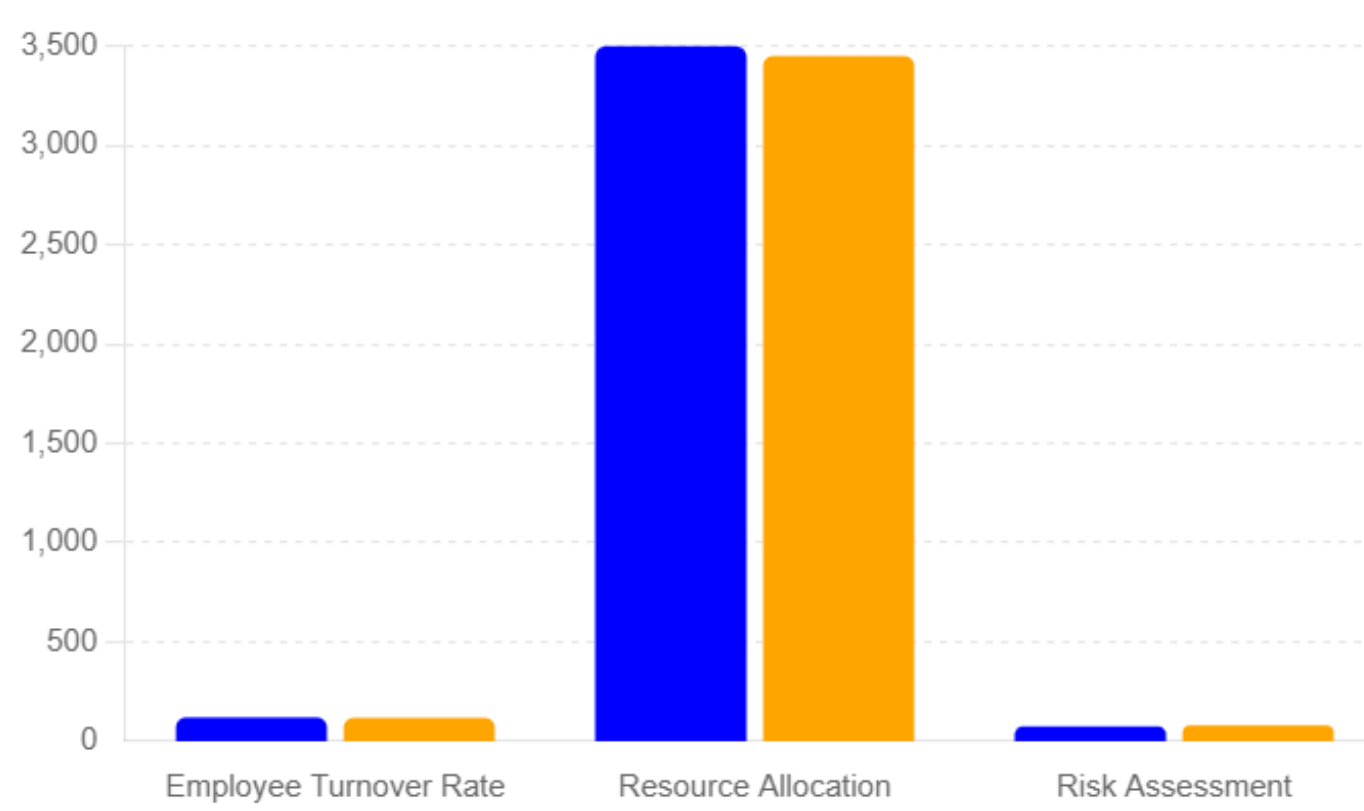

Fig. 2. Predictive accuracy: Predicted vs. Actual Values

Additionally, TalentOptima's focus on predictive accuracy and ability to automate routine tasks distinguishes it from competitors like Trello and ClickUp, which are more geared toward more straightforward project management needs. Including advanced features such as integrated budget tracking and financial performance metrics in TalentOptima also provides a more comprehensive solution than tools like Smartsheet and Basecamp, which lack built-in cost management functionalities. This holistic approach makes TalentOptima particularly well suited for complex projects where precise planning and efficient resource utilization are essential. The table shows these distinctions, showing that many tools offer robust foundational features. However, TalentOptima's advanced capabilities provide a significant edge in driving project success through better decision making and strategic foresight.

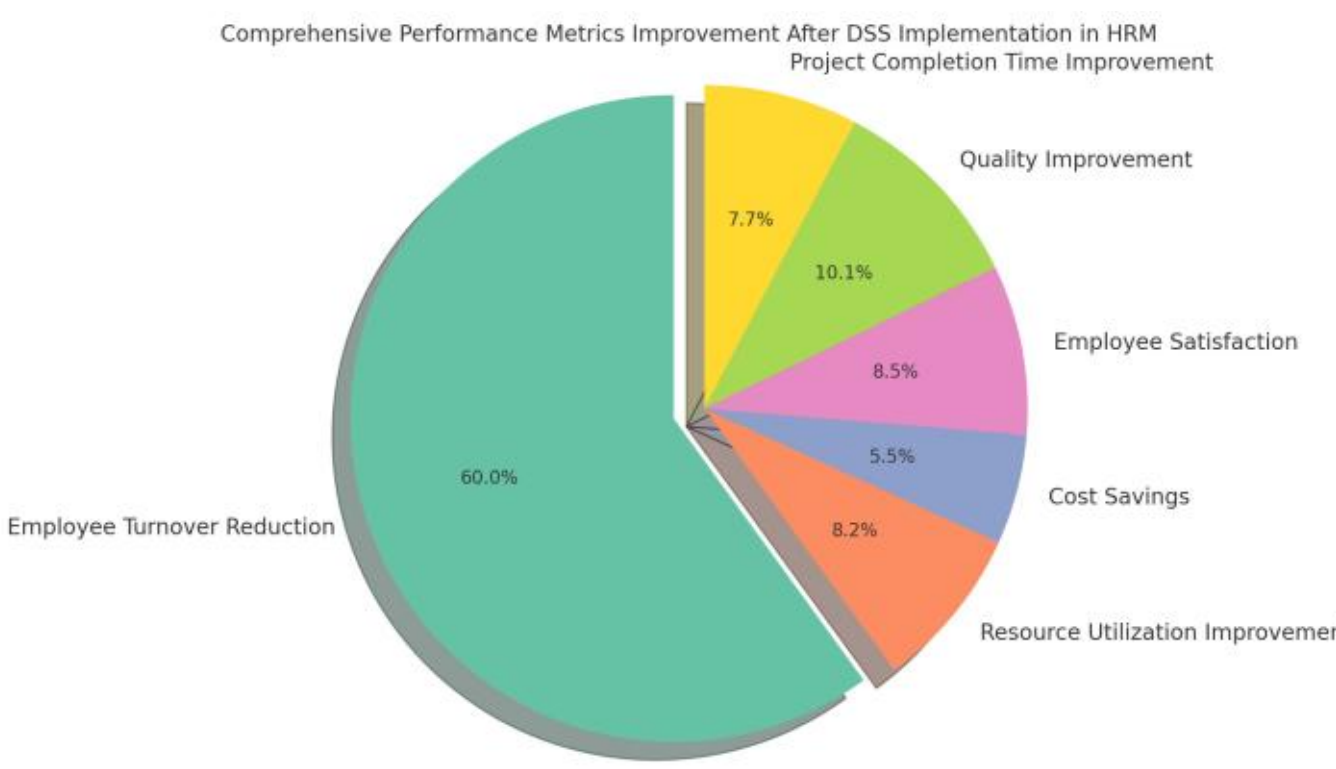

Fig. 3. Improvement of comprehensive performance metrics after implementation of DSS mentation in HRM

## VI. Conclusion

The evaluation of TalentOptima demonstrates its significant impact in improving software project management through its advanced predictive analytics, AI-driven decision making, and dynamic resource management capabilities. The outcomes indicate a marked improvement in project completion times, resource utilization and risk mitigation, as reflected in quantitative performance metrics and user feedback. TalentOptima offers a more integrated approach, seamlessly combining real-time data analysis, automated workflows, and comprehensive budget and cost management features compared to other leading project management tools. These technical advantages streamline project processes and provide project managers with deeper insights, enabling more informed decisions and better overall project outcomes. TalentOptima's superior performance in these critical areas shows its value as a cutting-edge tool designed to meet the complex demands of modern software project management, positioning it as a leader in the field.

TABLE III Comparison of TalentOptima with Other HRM Tools

| Feature | TalentOptima | BambooHR [34] | Workday [35] | ADP [36] | Zenefits [37] | That is, [38] | Gusto [39] | SAP SuccessFactors [40] |
|---|---|---|---|---|---|---|---|---|
| Employee Management | AI-driven talent Optimization, Predictive analytics | Employee database, customizable workflows | Full-suite HR management, real-time data | Employee records, custom dashboards | Centralized Employee Management | HR database, customizable workflows | Employee database, benefits tracking | Comprehensive employee records, global workforce management |
| Recruitment and Onboarding | Predictive recruitment analytics, AI-assisted onboarding | Applicant tracking and onboarding workflows | Recruitment and talent acquisition modules. | Hiring, recruitment, onboarding | Applicant tracking, benefits, onboarding | Onboarding workflows, applicant tracking | Automated hiring, onboarding | Recruitment, onboarding, interview management |

| Employee Performance Tracking | Real-time performance tracking, AI-based assessments | Performance reviews, feedback system | Continuous performance management, real-time insights | Performance tracking, productivity reports | Goal Setting, Employee Feedback | Employee Performance Reviews | Performance tracking, productivity insights | Performance reviews, goal setting, continuous feedback |
|---|---|---|---|---|---|---|---|---|
| Time and At tendance | Integrated time tracking, AI-driven absence prediction | Time tracking, absence management | Attendance management, time-off requests | Time and Attendance tracking, scheduling | Time tracking, time-off management | Timesheets, Time-off tracking | Automated Time Tracking | Attendance, leave management, timesheets |
| Payroll and Compensation | Predictive compensation analysis, automated payroll | Payroll Processing, Compensation Management | Payroll management, compensation planning | Automated payroll, tax filings | Compensation tracking, payroll processing | Payroll processing, benefits management | Payroll, benefits, tax filings | Compensation, payroll, incentive management. |
| Benefits Ad-ministration | AI-optimized benefits allocation, cost management | Benefits tracking, Health Plan Management | Employee benefits, compensation packages, | Benefits ad-ministration, compliance | Benefits enrollment, health insurance | Benefits management, enrollment tracking | Benefits management, health plan Administration | Benefits management, global compliance |
| Compliance and Reporting | Real-time compliance monitoring, predictive reporting | HR reporting, compliance tracking | Compliance management, real-time reporting | Compliance and tax reporting | Compliance automation and reporting tools | Compliance Tracking, Reporting Dashboards | Compliance management, payroll reports | Global compliance management, real-time HR reporting |
| Employee Engagement | AI-driven engagement insights, feedback analysis | Employee feedback and engagement surveys | Employee engagement an-alytics | Feedback surveys, engagement tracking | Employee engagement tools, feedback loops | Employee Engagement, Surveys | Engagement surveys, feedback tracking | Engagement of employees insights, performance feedback |
| Integration and Automation | Seamless integration with HR tools, automation of recruitment, and payroll | Integration with third-party apps, automation workflows | Integration with various HR tools, full automation | Integration with HR tools, payroll automation | Automation of benefits, payroll, and recruitment | Integration with the other tools, workflow automation | Automatization of payroll, benefits | Integration with enterprise HR tools, full HR automation |
| Security and Permissions | Advanced access control, role-based security | Role-based permissions, secure data storage | Secure access control, role-based security | Secure access, customizable permissions | Data security, role-based access control | Role-based access, customizable permissions | Secure access, role management | Advanced security, role-based data access |